\documentclass[9pt,twocolumn,twoside]{osajnl}

\journal{ao-notitle} % Choose journal (ao, aop, josaa, josab, ol)
\usepackage{graphicx}
\usepackage{float,comment}
\usepackage[tight,footnotesize]{subfigure}
\usepackage{mathdots}
\usepackage{epstopdf}
\usepackage{hyperref}
\usepackage{xcolor}
\usepackage{amsmath,amssymb}
\usepackage{amsthm}
\usepackage{mathrsfs}
\DeclareMathOperator*{\argmax}{argmax}
\DeclareMathOperator*{\argmin}{argmin}

\setboolean{shortarticle}{false}
\ifthenelse{\boolean{shortarticle}}{\colorlet{color2}{color2b}}{\colorlet{color2}{color2}} % Automatically switches colors for short articles

\title{Efficient illumination angle self-calibration in Fourier ptychography}

\author[1,*]{Regina Eckert}
\author[2]{Zachary F. Phillips}
\author[1]{Laura Waller}

\affil[1]{Department of Electrical Engineering and Computer Sciences, University of California, Berkeley, California 94720, USA}
\affil[2]{Graduate Group in Applied Science and Technology, University of California, Berkeley, California 94720, USA}

\affil[*]{Corresponding author: eckert@berkeley.edu}

\dates{Compiled \today}

\ociscodes{(110.1758) Computational imaging; (100.5070) Phase retrieval; (110.2945) Illumination design}

\doi{\url{http://dx.doi.org/10.1364/ao.XX.XXXXXX}}

\begin{abstract}
Fourier ptychography captures intensity images with varying source patterns (illumination angles) in order to computationally reconstruct large space-bandwidth-product images. Accurate knowledge of the illumination angles is necessary for good image quality; hence, calibration methods are crucial, despite often being impractical or slow. Here, we propose a fast, robust, and accurate self-calibration algorithm that uses only experimentally-collected data and general knowledge of the illumination setup. First, our algorithm makes a direct estimate of the brightfield illumination angles based on image processing. Then, a more computationally-intensive spectral correlation method is used inside the iterative solver to further refine the angle estimates of both brightfield and darkfield images. We demonstrate our method for correcting large and small misalignment artifacts in both 2D and 3D Fourier ptychography with different source types: an LED array, a galvo-steered laser, and a high-NA quasi-dome LED illuminator. 

\end{abstract}

\setboolean{displaycopyright}{true}

\begin{document}

\maketitle
\thispagestyle{fancy}

\ifthenelse{\boolean{shortarticle}}{\ifthenelse{\boolean{singlecolumn}}{\abscontentformatted}{\abscontent}}{}

\section{Introduction}

Computational imaging leverages the power of both optical hardware and computational algorithms to reconstruct images from indirect measurements. In optical microscopy, programmable illumination sources have been used for computational illumination techniques including multi-contrast~\cite{Zheng2011,Liu2014}, quantitative phase~\cite{Zheng2013,Tian:14,Tian:15,Chen2016} and super-resolution~\cite{Zheng2013,Ou2013,Dong:14,Tian2014,Tian2015} microscopy. However, most of these methods are very sensitive to experimental misalignment errors and can suffer severe artifacts due to model mismatch. Extensive system calibration is needed to ensure the inverse algorithm is consistent with the experimental setup, which can be time- and labor-intensive. This often requires significant user expertise, which makes the setup less accessible to reproduction by non-experts and undermines the simplicity of the imaging scheme. Further, pre-calibration methods are not robust to changes in the system (e.g. bumping the setup, changing objectives, sample-induced aberrations) and often require very precise ground truth test objects.

Algorithmic self-calibration methods ~\cite{Thibault2009,Bian:13,Ou:14,Horstmeyer:14,Yeh2015,Bian:16,Chung:16fluor,Eckert:16,Sun:16LEDpos,Satat2016,Pan2017,Dou2017,Liu:17} eliminate the need for pre-calibration and ground truth test objects by making calibration part of the inverse problem. These methods jointly solve two inverse problems: one for the reconstructed image of the object and the other for the calibration parameters. By recovering system calibration information directly from captured data, the system becomes robust to system changes, such as misalignment or sample-induced aberrations.

\begin{figure*}[t]
	\centering
	\includegraphics[width=0.85\textwidth]{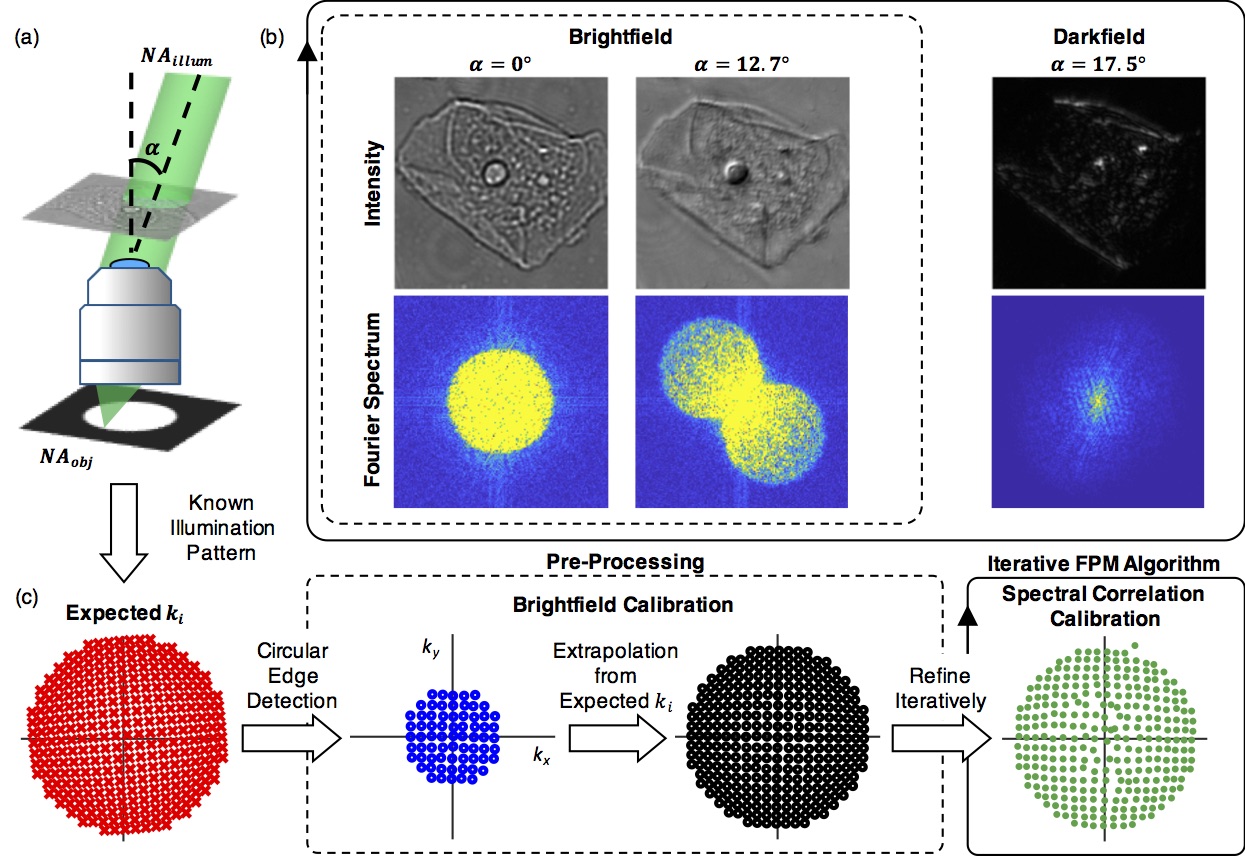}%.pdf}
	\caption{Illumination angles are calibrated by analyzing Fourier spectra in our method. (a) A cheek cell is illuminated at angle $\alpha$ and imaged with $NA_{obj}$. (b) Brightfield images contain overlapping circles in their Fourier spectra; darkfield images do not. (c) We perform a brightfield calibration in pre-processing, extrapolate the correction to darkfield images, then iteratively calibrate angles inside the FPM algorithm using spectral correlation calibration.
		\vspace{-5mm}}
	\label{fig:Overview}
\end{figure*}

Here, we focus on \textit{illumination angle} self-calibration for Fourier Ptychographic Microscopy (FPM)~\cite{Zheng2013}. FPM is a coherent computational imaging method that reconstructs high-resolution amplitude and phase across a wide field-of-view (FoV) from intensity images captured with a low-resolution objective lens and a dynamically-coded illumination source. Images captured with different illumination angles are combined computationally in an iterative phase retrieval algorithm that constrains the measured intensity in the image domain and pupil support in the Fourier domain. This algorithm can be described as stitching together different sections of Fourier space (synthetic aperture imaging~\cite{Turpin:1995,Di:08}) coupled with iterative phase retrieval. FPM has enabled fast \textit{in vitro} capture via multiplexing~\cite{Tian2014,Tian2015}, fluorescence imaging~\cite{Chung:16fluor}, and 3D microscopy~\cite{Tian20153D,Horstmeyer:16}. It requires at least 35\% overlap between adjacent angles of illumination~\cite{Dong:14,Sun:16}, providing significant redundancy in the dataset and making it suitable for joint estimation self-calibration formulations. 

Self-calibration routines have previously been developed to solve for pupil aberrations~\cite{Ou:14}, illumination angles~\cite{Yeh2015,Sun:16LEDpos,Liu:17}, LED intensity~\cite{Bian:13}, sample motion~\cite{Bian:16}, and auto-focusing~\cite{Dou2017} in FPM. The state-of-the-art self-calibration method for illumination angles is simulated annealing~\cite{Yeh2015,Sun:16LEDpos}, a joint estimation solution which under proper initialization successfully removes LED misalignment artifacts that usually manifest as low-frequency noise. Unfortunately, because the simulated annealing procedure optimizes illumination angles inside the FPM algorithm, it slows the solver by an order of magnitude or more, significantly increasing run-times. In fact, the computational costs become infeasible for 3D FPM (which is particularly sensitive to angle calibration~\cite{Tian20153D}). 

Moreover, most self-calibration algorithms require the calibration parameters to be initialized close to their optimum values. This is especially true when the problem is non-convex or if multiple calibration variables are to be solved for (\textit{e.g.} object, pupil, and angles of illumination). Of the relevant calibration variables for FPM, illumination angles are most prone to misestimation over time and across different illumination systems and samples, due to shifts or rotations of the LED array~\cite{Guo:15}, laser illumination instabilities~\cite{Kuang:15,Eckert:16}, difficult to predict non-planar illuminator arrangements~\cite{Phillips2015,Chung2016,Sen:16,Phillips:17}, or sample-induced aberrations~\cite{Hell:1993,Kang:18}. Samples can also change the effective illumination angles dynamically, such as when moving in an aqueous solution that bends the light unpredictably. Efficient angle self-calibration based on captured data is necessary to make FPM systems robust to these changes.

We propose here a two-pronged angle self-calibration method that uses both pre-processing (\textit{brightfield calibration}) and iterative joint estimation (\textit{spectral correlation calibration}) that is quicker and more robust to system changes than state-of-the-art angle calibration methods. A circle-finding step prior to the FPM solver accurately identifies the angles of illumination in the brightfield (BF) region. A transformation between the expected and BF calibrated angles extrapolates the correction to illuminations in the darkfield (DF) region. Then, a local grid-search-based algorithm inside the FPM solver further refines the angle estimates, with an optional prior based on the illuminator geometry to make the problem more well-posed. Our method is object-independent, robust to coherent noise, and time-efficient, adding only seconds to the processing time. We demonstrate practical on-line angle calibration for 2D and 3D FPM with 3 different source types: an LED array, a galvonometer-steered laser, and a high-NA (max $NA_{illum} = 0.98$) quasi-dome illuminator~\cite{Phillips:17}.

\section{Methods}
The image formation process for a thin sample under off-axis plane wave illumination can be described by:
\begin{equation}
I_i = |\mathscr{F}(\tilde{O}(\mathbf{k}-\mathbf{k}_i)\cdot\tilde{P}(\mathbf{k}))|^2 = |O(\mathbf{r})e^{-i2\pi \mathbf{k}_i\mathbf{r}} * P(\mathbf{r})|^2,
\label{eq:forwardModel}
\end{equation}

\noindent where $\mathbf{k}_i$ is the spatial frequency of the incident light, $\tilde{P}(\mathbf{k})$ is the low-pass system pupil function, $\tilde{O}(\mathbf{k})$ is the object Fourier spectrum, and $\mathscr{F}$ represents the 2D Fourier transformation operation. Intensity images are captured at the camera plane, corresponding to auto-correlation in the Fourier domain:

\begin{equation}
\tilde{I_i} = \mathscr{F}(|O(\mathbf{r})e^{-i2\pi \mathbf{k}_i\mathbf{r}} * P(\mathbf{r})|^2).
\label{eq:Fourier domain}
\end{equation}

\noindent In the brightfield region, where illumination angles are within $NA_{obj}$, $|\tilde{I_i}|$ will contain two distinct circles centered at $\mathbf{k}_i$ and $-\mathbf{k}_i$ with radius $R = \frac{NA_{obj}}{\lambda}$ due to the strong interference of the pupil with the sample's zero-order (DC) term (see Fig.~\ref{fig:Overview}b). We calibrate in brightfield by finding the circle location. Darkfield images do not result in clearly defined circles in $|\tilde{I_i}|$ as the illumination's DC term is outside $NA_{obj}$. Therefore, we also iteratively calibrate inside the FPM solver by correlating overlapping spectra to calibrate angles relative to each other.

Both parts of our algorithm rely on analysis of each image's Fourier transform to recover illumination angles. Fourier domain pupil analysis of intensity images has been used previously to deduce imaging system aberrations~\cite{Shanker:16} and determine the center of diffraction patterns~\cite{DAMMER1997214,CAUCHIE2008567} for system calibration. We show here that the individual Fourier spectra can be used to determine illumination angle.

\subsection{Brightfield Calibration}
Locating the center of the circles in the amplitude of a Fourier spectrum is an image processing problem. Previous work in finding circles in images uses the Hough transform, which relies on an accurate edge detector as an initial step~\cite{Yuen89,Davies:04}. In practice, however, we found that edge detectors do not function well on our datasets due to speckle noise, making the Hough transform an unreliable tool for our purpose. Therefore, we propose a new method which we call \textit{circular edge detection}.

\begin{figure} [tb]
	\centering
	\includegraphics[width=0.48\textwidth]{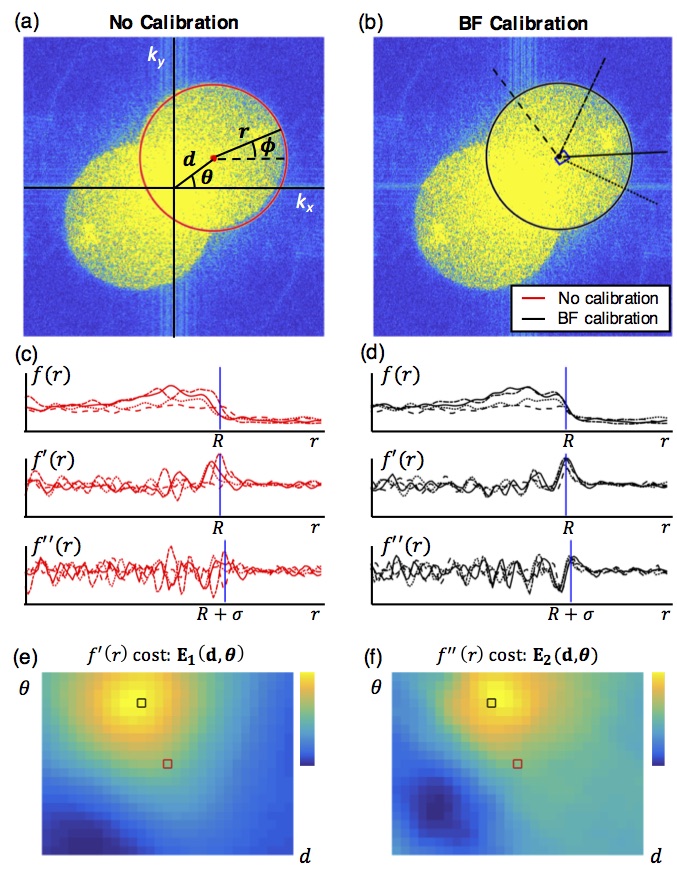}%.pdf}
	\caption{Circular edge detection on brightfield images finds circle centers, giving illumination angle. (a,b) Comparison of uncalibrated (red) and calibrated (black) illumination $\mathbf{k}_i$. The blue box in (b) indicates the search range for $\mathbf{k}_i$. (c,d) $\tilde{I_i}$ along radial lines, $f(r,\phi_n)$, and derivatives with respect to $r$. (h,i) $E_1$ and $E_2$, sums of the derivatives at known radii $R$ and $R+\sigma$, peak near the correct center. Boxes show uncalibrated (red) and calibrated (black) $\mathbf{k}_i$ centers.
		\vspace{-5mm}}
	\label{fig:BF_calibration}
\end{figure}

Intuitively, circular edge detection can be understood as performing edge detection (\textit{i.e.} calculating image gradients) along a circular arc around a candidate center point in k-space (the Fourier domain). To a first approximation, we assume $|\tilde{I_i}|$ is binary function that is 1 inside the overlapping circles and 0 everywhere else. Our goal is to find the strong binary edge in order to locate the circle center.  Based on information we have about our illumination set-up, we \textit{expect} the illumination spatial frequency (and therefore circle center) for spectrum $\tilde{I_i}$ to be at $\mathbf{k}_{i,0} = (k_{x,i,0},k_{y,i,0})$ (polar coordinates $\mathbf{k}_{i,0} = (d_{i,0}, \theta_{i,0})$) (Fig.~\ref{fig:BF_calibration}a). If this is the \textit{correct} center  $\mathbf{k}_i'$, we expect there to be a sharp drop in $|\tilde{I_i}|$ at radius $R$ along any radial line $f(r,\phi_n)$ out from $\mathbf{k}_i'$ (Fig.~\ref{fig:BF_calibration}b,d). This amplitude edge will appear as a peak at $r=R$ in the first derivative of each radial line with respect to $r$, $f'(r,\phi_n)$. Here $(r,\phi_n)$ are the polar coordinates of the radial line with respect to the center $\mathbf{k}_i$, considering the $n^{th}$ of $N$ total radial lines. 

We identify the correct $\mathbf{k}_i'$ by evaluating the summation of the first derivative around the circular arc at $r=R$ from several candidate $\mathbf{k}_i = (d_i,\theta_i)$:

\begin{equation}
E_1(R, d_i,\theta_i)=\sum_{n=1}^{N} f'(r = R,\phi_n,d_i,\theta_i).
\label{eq:eprime}
\end{equation}

 \noindent When $\mathbf{k}_i$ is incorrect, the edges do not align and the derivative peaks do not add constructively at $R$ (Fig.~\ref{fig:BF_calibration}c). The derivatives at $R$ are all maximized \textit{only} at the correct center $\mathbf{k}_i'$ (Fig.~\ref{fig:BF_calibration}d), creating a peak in $E_1$ (Fig.~\ref{fig:BF_calibration}e). This is analogous to applying a classical edge filter in the radial direction from a candidate center and accumulating the gradient values at radius $R$.

In order to bring our data closer to our binary image approximation, we divide out the average spectrum $|\tilde{I_i}|$ across all $i$ spectra, essentially removing the effect of the object $\tilde{O}(\mathbf{k})$ from each spectrum. We then convolve with a Gaussian blur kernel with standard deviation $\sigma$ to remove speckle noise (Alg.~\ref{alg:BF}.\ref{alg:1}-\ref{alg:2}). Under this model, the radial line $f(r,\phi_n)$ from our correct center $\mathbf{k}_i'$ is a binary step function convolved with a Gaussian:

\begin{equation}
f(r,\phi_n,d_i',\theta_i') = \text{rect}(\frac{r}{2R}) * \frac{1}{\sqrt{2\pi}\sigma}e^{\frac{-r^2}{2\sigma^2}}.
\label{eq:spoke}
\end{equation}

\noindent By differentiating, we find the peak of $f'(r,\phi_n)$ still occurs at $r=R$. Additionally, we find that the second derivative $f''(r,\phi_n)$ is maximized at $r=R+\sigma$. Experimentally, we have found that considering both the first \textit{and} second derivatives increases our accuracy and robustness to noise across a wide variety of datasets. We therefore calculate a second derivative metric,

\begin{equation}
E_2(R+\sigma,d_i,\theta_i)=\sum_{n=1}^{N} f''(r = R+\sigma,\phi_n,d_i,\theta_i),
\label{eq:edoubleprime}
\end{equation}

\noindent which is jointly considered with Eq.~\ref{eq:eprime}. We identify candidate centers $\mathbf{k}_i$ that occur near the peak of \textit{both} $E_1$ and $E_2$ (Fig.~\ref{fig:BF_calibration}e-f), then use a least-squares error metric to determine the final calibrated $\mathbf{k}_i'$ (Alg.~\ref{alg:BF}.\ref{alg:5}-\ref{alg:8}). In practice, we also only consider the non-overlapping portion of the circle's edge, bounding $\phi$.

\begin{figure*}[htb]
	\centering
	\includegraphics[width=1\textwidth]{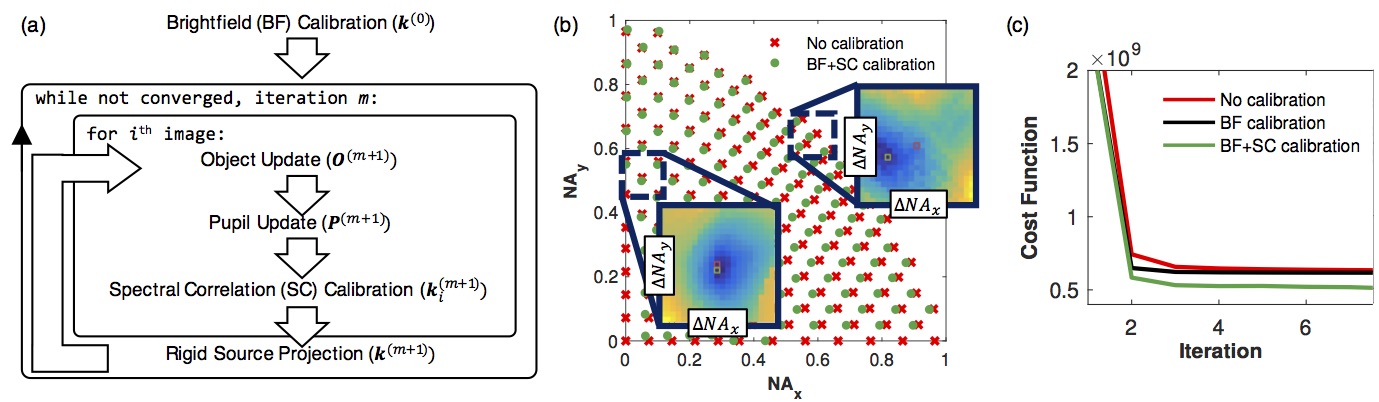}%.pdf}
	\caption{BF calibration pre-calibrates the illumination angles and SC calibration refines iteratively within FPM. (a) Algorithm block diagram. (b) Uncalibrated (red) and BF + SC calibrated (green) illumination angle map. Inset are example search spaces for optimal illumination angles, showing local convexity. (d) FPM convergence plot across methods.
		\vspace{-5mm}}
	\label{fig:DF_calibration}
\end{figure*}

Up to this point, we have assumed that the precise radius $R$ of the imaging pupil in $\tilde{I_i}$ is known. However, in pixel units, $R$ is dependent on the pixel size of the camera, $p_s$, and the system magnification, $mag$:

\begin{equation}
R = \frac{NA_{obj}}{\lambda} \frac{p_s*M}{mag}
\end{equation}

\noindent as well as $NA_{obj}$ and $\lambda$, where $\tilde{I_i}$ is dimension $MxM$. Given that $mag$ and $NA_{obj}$ are often imprecisely known but are unchanged across all images, we calibrate the radius by finding the $R'$ which gives the maximum gradient peak $E_1$ across multiple images before calibrating $\mathbf{k}_i'$ (Alg.~\ref{alg:BF}.\ref{alg:3}). A random subset of images is used to decrease computation time.

Finally, once all images are calibrated, we want to remove outliers and extrapolate the correction to the darkfield images. Outliers occur due to: 1) little high-frequency image content, and therefore no defined circular edge; 2) strong noise background; or 3) large shifts such that the conjugate circle center $-\mathbf{k}_i$ is identified as $\mathbf{k}_i'$. In these cases, we cannot recover the correct center based on a single image and must rely on the overall calibrated change in the illuminator's position. We find outliers based on an illuminator-specific transformation $A$ (\textit{e.g.}, rigid motion) between the expected initial guess of circle centers $\mathbf{k}_{i,0}$ (\textit{e.g.}, the LED array map) and the calibrated centers $\mathbf{k}_i'$ using a RANSAC-based method~\cite{Jacobson:15}. This transformation is used to correct outliers and darkfield images (Alg.~\ref{alg:BF}.\ref{alg:9}-\ref{alg:12}), serving as an initialization for our spectral correlation (SC) method.

\begin{algorithm}
\caption{Brightfield Calibration}\label{alg:BF}
\begin{algorithmic}[1]
\State $\tilde{I}_f \gets |\tilde{I}|/\textrm{mean}_i(|\tilde{I_i}|)$\Comment{Divide out mean spectrum}\label{alg:1}
\State $\tilde{I}_f \gets \textrm{gauss}(\tilde{I}_f,\sigma)$\Comment{Smooth speckle}\label{alg:2}
\State $R' \gets \argmax_R E_1(R,d_i,\theta_i), \textrm{subset }(\tilde{I}_{f,i})$\Comment{Calibrate radius}\label{alg:3}
\For{$i^{th}$ image}
\Comment{Circular edge detection}
\State $\mathbf{k}_{i,1} \gets (d_i,\theta_i) \textrm{ where } E_1 \textrm{ near max (within 0.1 std)}$\label{alg:5}
\State $\mathbf{k}_{i,2} \gets (d_i,\theta_i) \textrm{ where } E_2 \textrm{ near max}$
\State $\mathbf{k}_{i} \gets \mathbf{k}_{i,1} \cap \mathbf{k}_{i,2}$\Comment{Consider both metrics}
\State $\mathbf{k}_i' \gets \argmin_{\mathbf{k}_i} ||I_i - \mathscr{F}(\tilde{I_i}\cdot\tilde{P}(\mathbf{k}-\mathbf{k_i}))||_2$ \Comment{Evaluate $\mathbf{k}_i$}
\EndFor \label{alg:8}
\State $A, i_{outliers} \gets \textrm{RANSAC}(A=\mathbf{k}_i'\backslash\mathbf{k}_{i,0})$\Comment{Identify outliers} \label{alg:9}
\State $\mathbf{k}_{inliers}^{(0)} \gets \mathbf{k}_{inliers}'$\Comment{Initialize for FPM}
\State $\mathbf{k}_{outliers}^{(0)} \gets A\mathbf{k}_{outliers,0}$
\State $\mathbf{k}_{darkfield}^{(0)} \gets A\mathbf{k}_{darkfield,0}$ \label{alg:12}

\end{algorithmic}
\end{algorithm}

\subsection{Spectral Correlation Calibration}
While the brightfield (BF) calibration method localizes illumination angles using intrinsic contrast from each measurement, this contrast is not present in high-angle (darkfield) measurements (Fig.~\ref{fig:Overview}b). Therefore, it is necessary to cast the angle calibration problem as a more general alternating minimization, where the object $O(\mathbf{r})$, pupil $P(\mathbf{k})$, and illumination angles $\mathbf{k_i}$ are jointly optimized at each iteration of the existing FPM algorithm. At the $m^{th}$ FPM iteration, we estimate the $i^{th}$ illumination angle by correlating spectrum $\tilde{I_i}$ with the updated object solution $O^{(m+1)}$. This correlation finds the relative k-space location of the current spectrum $\tilde{I_i}$ relative to the overall object, giving an estimate $\mathbf{k}_i^{(m)}$ relative to the other illuminator angles $\mathbf{k}^{(m)}$. In practice, we do not know the correct illumination wave vector $\mathbf{{k}}_i'$, and must search for the $\Delta\mathbf{{k}}$ that maps our guess $\mathbf{{k}}_i^{(m)}$ to $\mathbf{{k}}_i'$. 

In order to minimize the forward model (Eq.~\ref{eq:forwardModel}) with respect to $\mathbf{k}_i=(k_{x,i},k_{y,i})$, we perform a discrete local grid search across adjacent illumination angles within each sub-iteration of the FPM algorithm. Our k-space resolution is band-limited by the field of view (FoV) of the object, which in FPM generally consists of small patches where the illumination can be assumed coherent. This FoV imposes a minimum resolvable discretization of illumination angles $\Delta{k} = \frac{2}{FoV}$ in k-space due to the Nyquist criterion. Since we cannot resolve angle changes beneath this resolution, we need only perform a local grid search over integer steps of $\Delta{k}$, which can be implemented with minimal computation of the forward model. Paired with the good initialization provided by our BF calibration, this makes our joint estimation SC method much faster than previous methods. 

Our problem is cast as an optimization of $\mathbf{k}_i^{(m+1)} = \mathbf{{k}}_i^{(m)} +\mathbf{n} \mathbf{\Delta {k}}$, where $\mathbf{n}\in [-1, 0 ,1]$ is the integer perturbation by $\Delta {k}$ of the current illumination angle $\mathbf{{k}}_i^{(m)}$.  We search all 9 values of $\mathbf{n}$ and choose the shift which leads to the smallest cost: 

\begin{equation}\label{Eq:positionOpt}
    \begin{aligned}
    & \underset{\mathbf{n}}{\text{argmin}}
    & & ||I_{i}- |O^{(m+1)} e^{-i2\pi (\mathbf{{k}}_i^{(m)} + \mathbf{n} \Delta\mathbf{{k}})\mathbf{\vec{r}}} * P^{(m+1)}|^2||_2^2 \\
    & \text{subject to}
    & &  \mathbf{n} \in [-1, 0 ,1]
    \end{aligned}
\end{equation}

\noindent This grid search is performed iteratively within each sequential iteration of an FPM reconstruction until $\mathbf{k}_i$ converges.

\begin{figure*} [htb]
	\centering
	\includegraphics[width=1\textwidth]{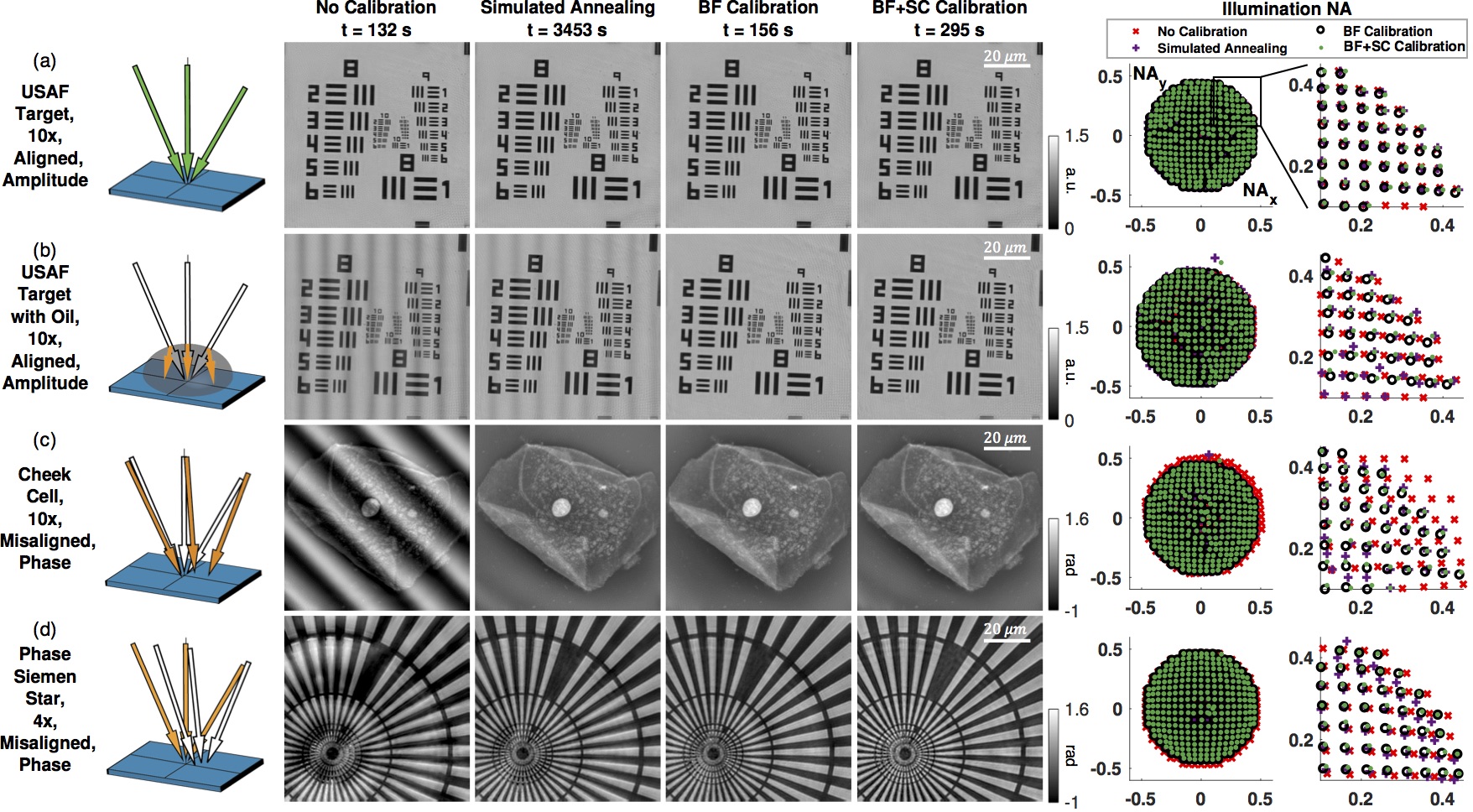}%.pdf}
	\caption{Conventional LED array microscope experimental results, comparing reconstructions with no calibration (average reconstruction time 132 seconds), simulated annealing (3453 s), our BF calibration (156 s), and our BF + SC calibration (295 s). (a) Amplitude reconstructions of a USAF target in a well-aligned system. (b) Amplitude reconstructions of the same USAF target with a drop of oil placed on top of the sample, distorting illumination angles. (c) Phase reconstructions of a human cheek cell with computationally misaligned illumination. (d) Phase reconstructions of a Siemens star phase target with experimentally misaligned illumination.
		\vspace{-5mm}}
	\label{Fig:results}
\end{figure*}

Including prior information about the illumination source makes our calibration problem more well-posed. For example, we know that an LED array is a rigid, planar illuminator. This knowledge about the illuminator is encoded in our initial expected illumination angle map, $\mathbf{k}_{i,0}$. By forcing our current estimates $\mathbf{k}_i^{(m)}$ to fit a transformation of this initial angle map at the end of each FPM sub-iteration, we can use this knowledge to regularize our optimization. The transformation model used also depends on the specific illuminator. For our quasi-dome LED array, which is composed of five circuit boards with precise LED positioning within each board, imposing an affine transformation from the angle map of each board to the current estimates $\mathbf{k}_i^{(m)}$ significantly reduces the problem dimensionality and mitigates noise across LEDs, making the reconstruction more stable.

\section{Results}

\subsection{Planar LED Array}

We first show experimental results from a conventional LED array illumination system with a 10x, 0.25 NA and a 4x, 0.1 NA objective lens at $\lambda = 514 nm$ and $NA_{illum} \leq 0.455 NA$. We compare reconstructions with simulated annealing, our BF pre-processing alone, and our combined BF+SC calibration method. All methods were run in conjunction with EPRY pupil reconstruction. We include results with and without the SC calibration to illustrate that the BF calibration is sufficient to correct for most misalignment of the LED array since we can accurately extrapolate LED positions to the darkfield region when the LEDs fall on a planar grid. However, when using a low NA objective ($NA_{obj} \leq 0.1$), as in Fig.~\ref{Fig:results}d, the SC method becomes necessary because the BF calibration is only able to use 9 images (compared to 69 brightfield images with a 10x, 0.25 NA objective, as in Fig.~\ref{Fig:results}a-c). In practice, a combination of our two methods ensures that LED positions will be accurately recovered.

Our method is object-independent, working for phase and amplitude targets as well as biological samples. All methods reconstruct similar quality results for the well-aligned LED array with the USAF resolution target. To simulate an aqueous sample, we place a drop of oil on top of the resolution target. The drop causes uneven changes in the illumination, giving low-frequency artifacts in the uncalibrated and simulated annealing cases which are corrected by our method. Our method is also able to recover a $5^{\circ}$ rotation, 0.02 NA shift, and 1.1x scaled computationally-imposed misalignment on well-aligned LED array data for a cheek cell, and gives a good reconstruction of an experimentally misaligned LED array for a phase Siemens star (Benchmark Technologies, Inc.). In contrast to simulated annealing, which on average takes $26 \times$ as long to process as FPM without calibration, our brightfield calibration only takes an additional 24 seconds of processing time and the combined calibration takes roughly only $2.25 \times$ as long as no calibration.

\subsection{Steered Laser}

\begin{figure*} [t]
	\centering
	\includegraphics[width=1\textwidth]{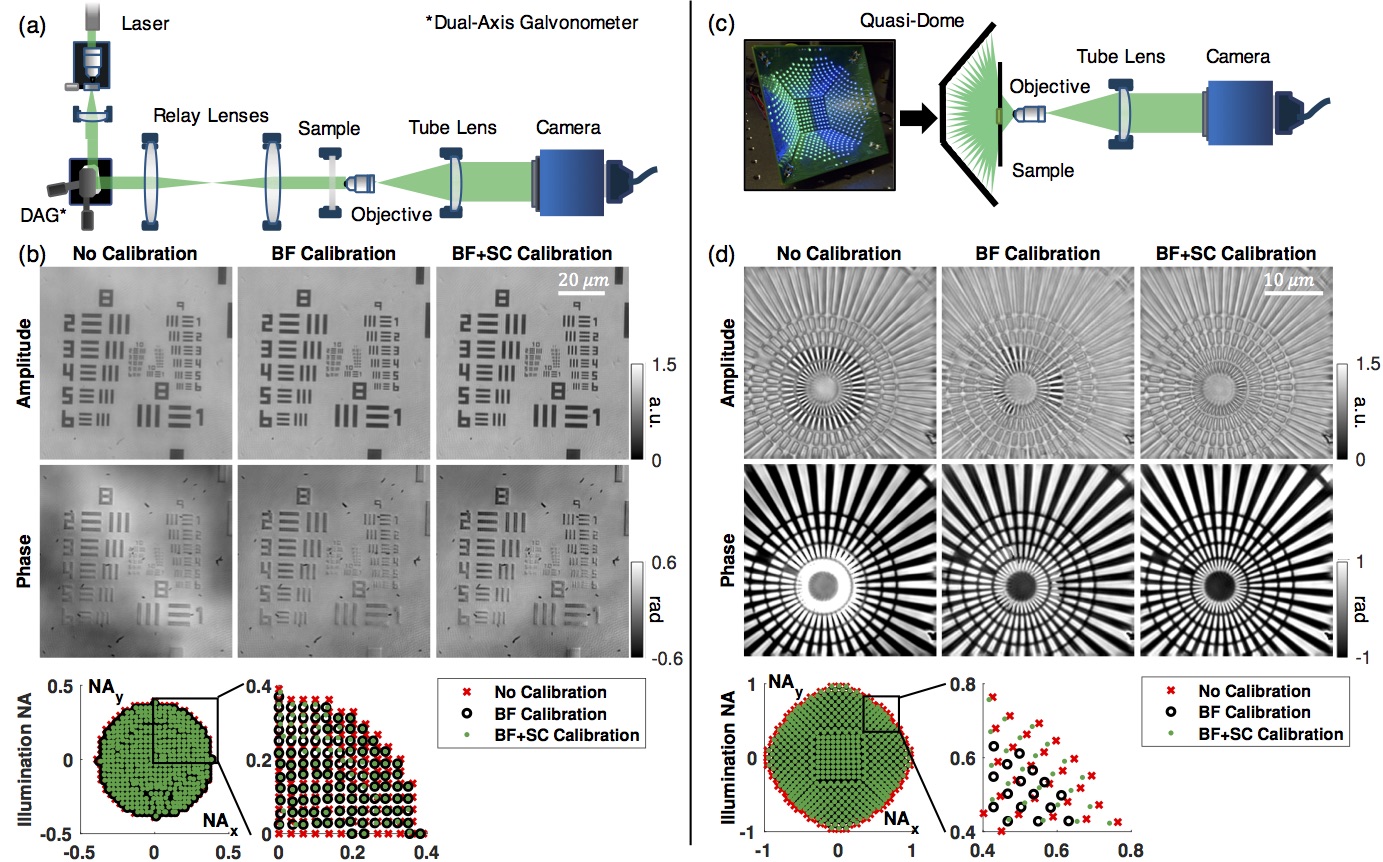}%.pdf}
	\caption{Experimental angle calibration in laser and high-NA quasi-dome illumination systems. (a) Laser illumination is steered by a dual-axis galvonometer. The angled beam is relayed to the sample by 4", 80 mm focal length lenses. Our calibration method removes low-frequency reconstruction artifacts. (b) The quasi-dome illuminator enables up to 0.98 $NA_{illum}$ using programmable LEDs. Our 1.23 NA reconstruction provides isotropic 425 $nm$ resolution with BF + SC calibration.
		\vspace{-5mm}}
	\label{Fig:laserDome}
\end{figure*}

Recent works have proposed laser illumination systems to increase the coherence and light efficiency of FPM~\cite{Kuang:15,Chung2016}. In practice, these systems are generally less rigidly aligned than LED arrays, making them more difficult to calibrate. To verify the performance of our method on these systems, we constructed a laser-based FPM system using a dual-axis galvonometer to steer a 532 $nm$, 5 mW laser, which is focused on the sample by large condenser lenses (Fig.~\ref{Fig:laserDome}). This laser illumination system allows finer, more agile illumination control than an LED array, as well as higher light throughput. However, the laser illumination angle varies from the expected value due to offsets in the dual-axis galvonometer mirrors, relay lens aberrations, and mirror position misestimations when run at high speeds. We can correct for these problems using our angle calibration method in a fraction of the time of previous methods.

\subsection{Quasi-Dome}

Since FPM resolution is defined by $NA_{obj} + NA_{illum}$, there is clear motivation to create high-NA illuminators for large space-bandwith product FPM reconstructions ~\cite{Sun2017,Phillips:17}. In order to illuminate at higher angles, the illuminators must become more dome-like, rather than planar, to maintain a good signal-to-noise ratio in the darkfield region~\cite{Phillips2015}. To address this, we previously developed a novel programmable quasi-dome array made of five separate planar LED arrays that can illuminate up to 0.98 NA~\cite{Phillips:17}. This device uses discrete LED control with RGB emitters ($\bar{\lambda}=[475nm, 530nm, 630nm]$) and can be attached to most commercial inverted microscopes.

As with conventional LED arrays, we assume that the LEDs on each board are rigidly placed as designed. However, each circuit board may have some relative shift, tilt, or rotation, since the final mating of the 5 boards is performed by hand. LEDs with high-angle incidence are both harder to calibrate and more likely to suffer from misestimation due to the dome geometry, so the theoretical reconstruction NA would be nearly impossible to reach without self-calibration. Using our method, we obtain the theoretical resolution limit available to the quasi-dome (Fig.~\ref{Fig:laserDome}). We show here that the SC calibration is especially important in the quasi-dome case, since it usually has many darkfield LEDs.

\subsection{3D FPM}

\begin{figure} [t]
	\centering
	\includegraphics[width=0.5\textwidth]{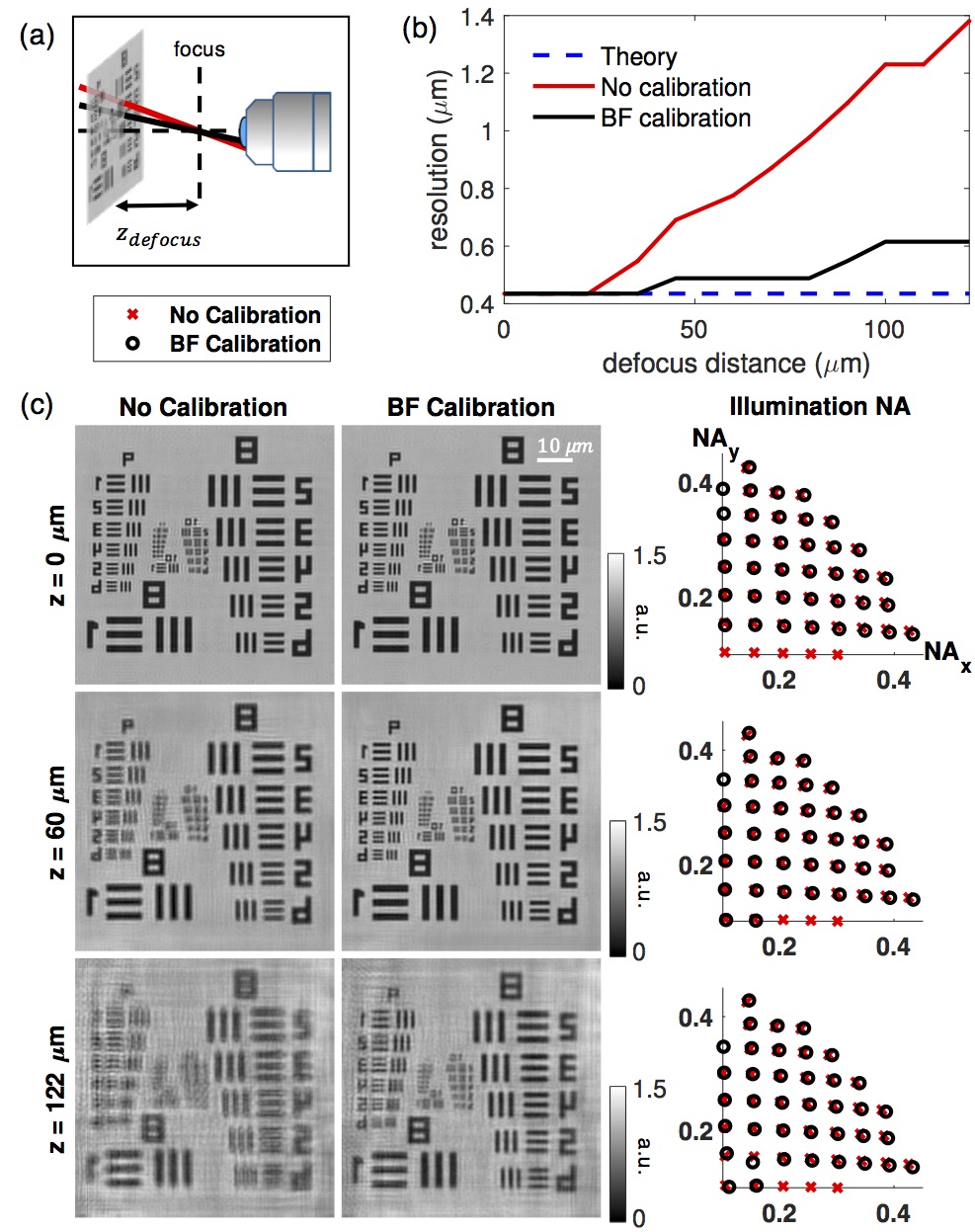}%.pdf}
	\caption{Even small calibration errors degrade 3D FPM resolution severely when defocus distances are large; our calibration algorithm reduces this error significantly. (a) Experiment schematic for a USAF target placed at varying defocus distances. (b) Measured reconstruction resolution. (c) Amplitude reconstructions for selected experimental defocus distances.  
		\vspace{-5mm}}
	\label{Fig:3D}
\end{figure}

Calibration is especially important for 3D FPM. Even small changes in angle become large when they are propagated to different depths, leading to reduced resolution and reconstruction artifacts~\cite{Tian20153D,Eckert:16}. For example, using a well-aligned LED array,~\cite{Tian20153D} was unable to reconstruct a resolution target defocused beyond 20 ${\mu}m$ at the theoretically possible resolution of 435 $nm$ due to angle misestimation; using the same dataset, our method allows us to reconstruct high-resolution features of the target even when it is 70 ${\mu}m$ off-focus (Fig.~\ref{Fig:3D}).

Since iterative angle joint-estimation, including our SC calibration, infeasibly increases the computational complexity of 3D FPM, we use BF calibration pre-processing only. While we do not attain the theoretical limits for all defocus depths, we offer significant reconstruction improvement. Fig.~\ref{Fig:3D}c reveals that our calibration only slightly changes the angles of illumination, highlighting that small angular changes have a large effect on 3D reconstructions. Experimental resolution was determined by resolvable bars on the USAF resolution target, where we declare a feature as "resolved" when there is a >20\% dip between $I_{max}$ and $I_{min}$.

\section{Discussion}
\begin{figure} [t]
	\centering
	\includegraphics[width=0.5\textwidth]{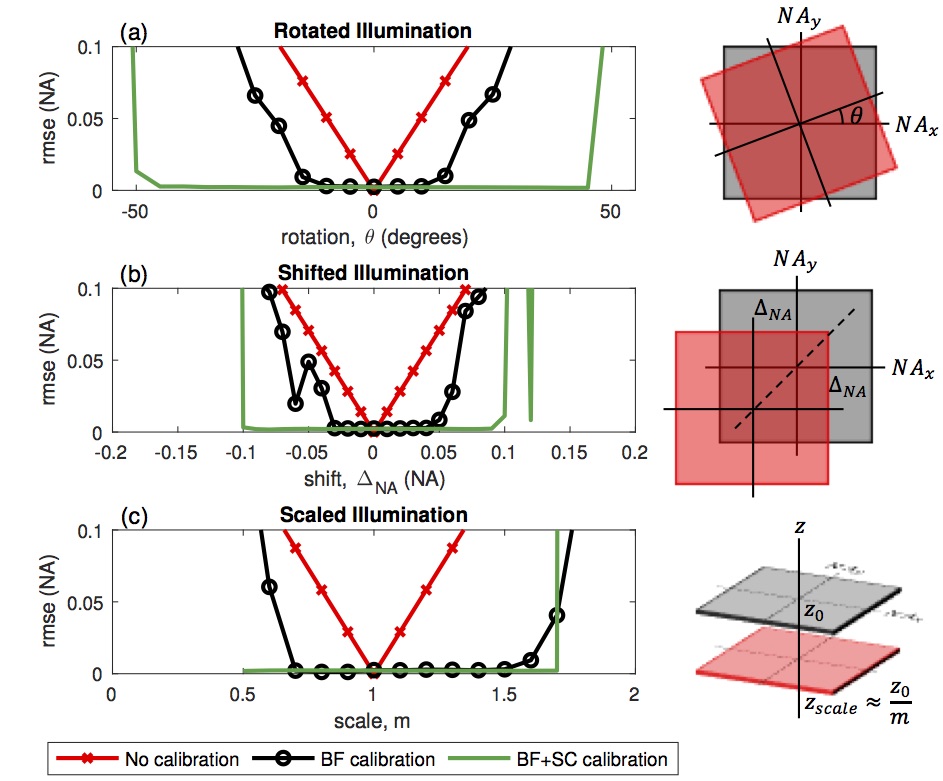}%.pdf}
	\caption{Our calibration is robust to large mismatches between estimated and actual LED array position. Simulation of misaligned illumination by (a) rotation, (b) shift, and (c) scale. Our calibration recovers the illumination with <0.005 NA error for rotations of $-45^{\circ}$ to $45^{\circ}$; shifts of -0.1 to 0.1 NA; and scalings of 0.5x to 1.75x before diverging.
		\vspace{-5mm}}
	\label{Fig:accuracy}
\end{figure}

Our calibration method offers significant gains in speed and robustness to large illumination misalignments over previous methods. BF calibration enables these capabilities by obtaining a good calibration that needs to be calculated only once in pre-processing, reducing computation. Since an estimation of a global shift in the illuminator based only on the brightfield images provides such a close initialization for the rest of the illumination angles, we can use a quicker, easier joint estimation computation in our SC calibration than would be otherwise possible. Jointly, these two methods work together to create fast and accurate reconstructions.

3D FPM algorithms are slowed an untenable amount by iterative calibration methods, since they require the complicated 3D forward model to be calculated multiple times during each iteration. Combined with 3D FPM's reliance on precise illumination angles to obtain a good reconstruction, it has previously been difficult to obtain accurate reconstruction of large volumes with 3D FPM. However, BF calibration can be used in the same way for both 2D and 3D datasets. Since the calibration occurs outside the 3D FPM algorithm, we can now correct for the angle misestimations that have degraded reconstructions in the past, allowing this algorithm to be applied to much larger volumes.

We analyze the robustness of our method to illumination changes by simulating an object illuminated by a grid of LEDs with $NA_{illum}<0.41$, with LEDs spaced at $0.041 NA$ intervals. We define the system to have $\bar{\lambda} = 532 nm$, with a 10x, 0.25 NA objective, a 2x system magnification, and a camera with $6.5 {\mu}m$ pixels. While the actual illumination angles in the simulated data remain fixed, we perturb the expected angle of illumination in typical misalignment patterns for LED arrays: rotation, shift, and scale (analogous to LED array distance from sample). We then calibrate the unperturbed data with the perturbed expected angles of illumination as our initial guess.

Figure~\ref{Fig:accuracy} shows that our method recovers the actual illumination angles with error less than 0.005 NA for rotations of $-45^{\circ}$ to $45^{\circ}$; shifts of -0.1 to 0.1 NA, or approximately a displacement by +/- 2 LEDs; and scalings of 0.5x to 1.75x (or LED array height between 40-140 $cm$ if the actual LED array height is 70 $cm$). In these ranges, the average error is 0.0024 NA, less than k-space resolution of 0.0032 NA. Our calibrated angles are very close to the actual angles even when the input expected angles were extremely far off. This result demonstrates that our method is robust to most unexpected alterations in the illumination scheme.

\section{Conclusion}

We have presented a novel two-part calibration method for recovering the illumination angles of a coherent illumination system for Fourier ptychography. We have demonstrated how this self-calibrating method makes Fourier ptychographic microscopes more robust to system changes and aberrations introduced by the sample. The method also makes it possible to use high-angle illuminators, such as the quasi-dome, and non-rigid illuminators, such as laser-based systems, to their full potential. Our pre-processing brightfield calibration further enables 3D multislice Fourier ptychography to reconstruct high resolution features across larger volumes than previously possible. These gains were all made with minimal additional computation, especially when compared to current state-of-the-art angle calibration methods, such as simulated annealing. Efficient self-calibrating methods such as these are important to make computational imaging methods more robust and available for broad use in the future. Open source code is available at \texttt{www.laurawaller.com/opensource}.

%\section{References}
% Bibliography
\bibliography{PhaseSpace}

% Full bibliography added automatically for Optics Letters submissions
% Note that this extra page will not count against page length
\ifthenelse{\equal{\journalref}{ol}}{%
%\clearpage
%\bibliographyfullrefs{PhaseSpace}
}{}

\end{document}